\documentstyle[epsf,amssymb,twocolumn,aps,psfig]{revtex}
\begin{document}
\twocolumn[\hsize\textwidth\columnwidth\hsize\csname @twocolumnfalse\endcsname
\

\title{Muon Spin Relaxation Measurements in Na$_{x}$CoO$_{2}$ $\cdot$ $y$H$_{2}$O}
\author{A. Kanigel, $^{\ast}$A. Keren, L. Patlagan, K. B. Chashka, B. Fisher}
\address{Department of Physics, Technion-Israel Institute of Technology, Haifa 32000, Israel.}
\author{P. King}
\address{$^{\ast}$Rutherford Appleton Laboratory, Chilton Didcot, Oxfordshire OX11 0QX, U.K.}
\author{A. Amato}
\address{Paul Scherrer Institute, CH 5232 Villigen PSI, Switzerland}
\pacs{PACS number}
\date{\today}
\maketitle

\begin{abstract}
Using the transverse field muon spin relaxation technique we measure the
temperature dependence of the magnetic field penetration depth $\lambda$, in
the Na$_{x}$CoO$_{2}\cdot y$H$_{2}$O system. We find that $\lambda,$ which is
determined by superfluid density $n_{s}$ and the effective mass $m^{\ast}$, is
very small and on the edge of the TF-$\mu$SR sensitivity. Nevertheless, the
results indicate that the order parameter in this system has nodes and that it
obeys the Uemura relation. By comparing $\lambda$ with the normal state
electron density we conclude that $m^{\ast}$ of the superconductivity carrier
is 70 times larger than the mass of bare electrons.

\end{abstract}
\vspace{0.5cm}
]

The discovery of the new superconductor Na$_{x}$CoO$_{2}\cdot y$H$_{2}$O
caused excitement in the unconventional superconductivity community. Of main
interest are three questions: (I) what is the symmetry of order parameter and
does it have nodes? (II) what is the gender of this material; is it a relative
of the cuprates, the heavy fermions, metallic superconductors, or a class in
its own? and (III) does this superconductor satisfy the Uemura relation?
Regarding the first question, several works have given contradictory answers.
For example $^{59}$Co NMR/NQR measurements by Kobayashi \textit{et. al.}
\cite{Kobayashi} and Wakiby \textit{et. al.} \cite{Waki} suggest the existence
of a coherence peak indicating a complete gap over the Fermi surface. In
contrast, Fujimoto \textit{et. al.} \cite{Fujimoto} and Ishida \textit{et.
al.} \cite{Ishida} found no coherence peak, questioning the previous result.
Therefore, an additional and different experimental approach is required. A
possible approach is to measure the temperature dependence of the magnetic
field penetration depth $\lambda$. At low temperatures, $\lambda$ is sensitive
to low-lying excitations, and, in the case of a complete gap $\lambda
(T)-\lambda(0)$ should vary exponentially as a function of $T$. On the other
hand, nodes in the gap lead to a power-law dependence of this penetration
depth difference. A study of $\lambda$ can help addressing the other two
questions as well. For the third question, one of the most universal
correlations among the unconventional superconductors is the relation between
the transition temperature $T_{c}$ and the width of the transverse field muon
spin rotation (TF-$\mu$SR) line at low temperatures, $\sigma(0)\propto
\lambda^{-2}$. Uemura \textit{et. al.} \cite{UemuraPRL91} were able to show
that the same relation holds for the underdoped cuprates, the bismuthates,
Chevrel-phase and the organic superconductors. This relation has no
explanation in the frame-work of the BCS theory, and it is usually explained
in terms of phase coherence establishment in a theory of local fluctuations of
the order parameter \cite{kivelson}. It is interesting to know if Na$_{x}%
$CoO$_{2}\cdot y$H$_{2}$O also obeys this relation. The second question would
be addressed by the absolute value of $\sigma(0)$.

The aim of this work is to measure the temperature dependence of $\lambda$
with TF-$\mu$SR in Na$_{x}$CoO$_{2}\cdot y$H$_{2}$O. TF-$\mu$SR is a very
useful way to study superconductors in the mixed state. In this method $100\%$
spin polarized muons are implanted in the sample, which is cooled in a field
perpendicular to initial muon spin. Above $T_{c}$, where the external field
penetrates the sample uniformly, the second moment of the field distribution
at the muon stopping site $\left\langle \Delta B^{2}\right\rangle $ is
relatively small and determined only by fields produced by nuclear moments.
Consequently the muon spins rotate in a coherent way and weak depolarization
of the muon ensemble is observed. When the sample is cooled below $T_{c}$ a
flux lattice (FLL) is formed in the sample resulting in an inhomogeneous field
and a therefore a larger second moment at the muon site. This increase in
$\left\langle \Delta B^{2}\right\rangle $ leads, in turn, to a high muon spin
depolarization rate in the sample. The penetration depth is related to the
field distribution width by
\begin{equation}
\left\langle \Delta B^{2}\right\rangle =\left(  \frac{0.00371F\Phi_{0}^{2}%
}{\lambda_{\perp}^{4}}\right)  ^{1/2}\label{deltaB}%
\end{equation}
where $\lambda_{\perp}$ is the in-plane penetration depth, $\Phi_{0}$ is the
flux quanta, and $F\sim0.44$ for anisotropic compounds \cite{BrandtPRB88}.

\begin{figure}
[ptb]
\centerline{\epsfxsize=9.0cm \epsfbox{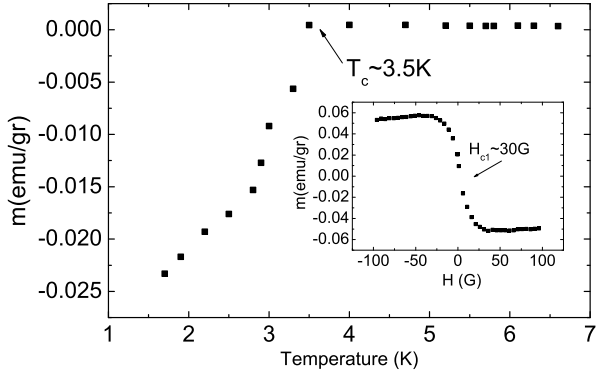}} 
\caption{Magnetization measurements of the Na$_{x}$CoO$_{2}\cdot y$H$_{2}$O vs
temperature. The transition temperature is $3.5$K. Inset: M vs. H for the same
sample. $H_{c1}\sim30~$G. }%
\label{mag}%
\end{figure}

Polycrystalline samples of Na$_{0.7}$CoO$_{2}$ were prepared by solid state
reaction \cite{Kawata} from mixtures of $Co$ and Na$_{2}$CoO$_{3}$. These
samples were intercalated as in Ref. \cite{Takada} using a solution of
Br$_{2}$ in CH$_{3}$CN, with Br$_{2}$ to Na molar ratio of $3$. Then the
material was washed in water and dried. The resulting compound was identified
as Na$_{0.3}$CoO$_{2}\cdot1.3$H$_{2}$O \cite{Fisher}. The transition
temperature was measured using a home-built DC magnetometer. In Fig.~\ref{mag}
we show the field cooled magnetization measured in a field of 50~Gauss. The
$T_{c}$ of the sample is about $3.5K$. In the inset we show the magnetization
vs. the applied field in zero field cooling conditions, measured at 1.8K. As
can be seen in the figure the lower critical field $H_{c1}$ is about 35~G.%

The $\mu$SR experiments were done on the MuSR spectrometer at the ISIS
spallation source at the Rutherford-Appelton Lab in the UK, and in the GPS
instrument on the $\pi$M3 beam line at PSI Switzerland. At ISIS we used the
$^{3}$He sorb cryostat which allows a base temperature of about $350$~mK. The
sample was cooled in a field of $400$~G, which is above $H_{c1}$. At PSI the
base temperature available was 1.6~K and we measured in field cooled
conditions in fields of $400~$G and $3~$kG. It has been reported that
Na$_{0.3}$CoO$_{2}\cdot1.3$H$_{2}$O is a very unstable system. It decays into
another hydrated phase which is non-superconducting. The system is sensitive
to temperature and humidity. For that reason we used a cell which allowed the
sample to be cooled without loosing the water content. The cell and the window
were made of Titanium, a material in which the muon polarization relaxes very
slowly. Although this material is superconducting below $400$~mK its $H_{c}$
is lower than our working field of $400$~G. We tried to minimize as much as
possible the exposure of the sample to dry and warm atmospheres. The sample
was exposed to lab atmosphere for no more than a few minutes before it was
sealed in the cell. The magnetization of the samples was measured to be the
same before and after the ISIS experiment.

\begin{figure}
[ptb]
\centerline{\epsfxsize=8.5cm \epsfbox{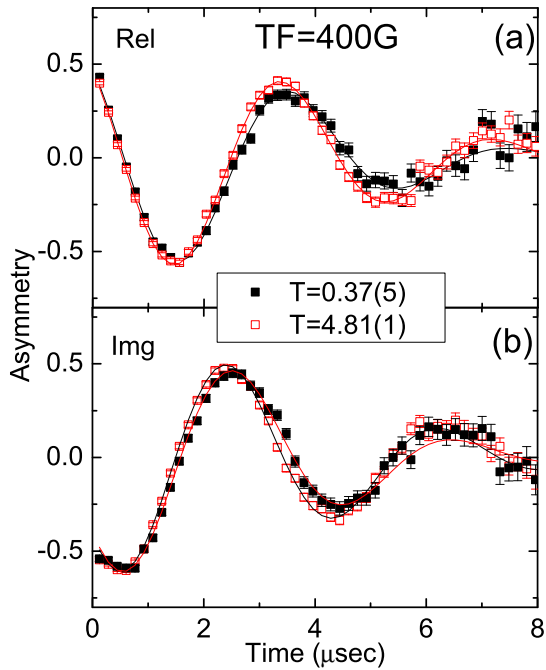}}
\caption{Transverse field $\mu$SR data at two different temperatures taken at
ISIS in 400~G. Both real and imaginary data are showed in a reference frame
rotating at $\gamma_{\mu}\times380$G. }%
\label{Raw}%
\end{figure}

The nuclear moment of the Co and the protons lead to relatively large
$\left\langle \Delta B^{2}\right\rangle $ and causes the muon polarization to
relax quite fast in Na$_{x}$CoO$_{2}\cdot y$H$_{2}$O even in the normal state.
Combined with fact that this compound might be an extreme type II SC with a
very large penetration depth, we expect only a small contribution to the
relaxation from the formation of the flux lattice. Usually, $\lambda\sim
10^{4}\mathring{A}$ is considered as the limit of the TF-$\mu$SR technique.
Here we expect values of that order, so very high statistics runs are needed.
In addition, the use of the ISIS facility which is optimized for weak
relaxation is an advantage. The MuSR spectrometer in ISIS consist of $32$
counters arranged on two circles. For demonstration purpose we combine all
$32$ counters using a RRF transformation and binning, and depict in Fig
\ref{Raw}(a) and (b) the imaginary and real rotation signals respectivly, at
both the highest and lowest temperatures. A small but clear difference is seen
in the relaxation rate between these two temperatures especially after 4~$\mu$sec.

\begin{figure}
[ptb]
\centerline{\epsfxsize=9.5cm \epsfbox{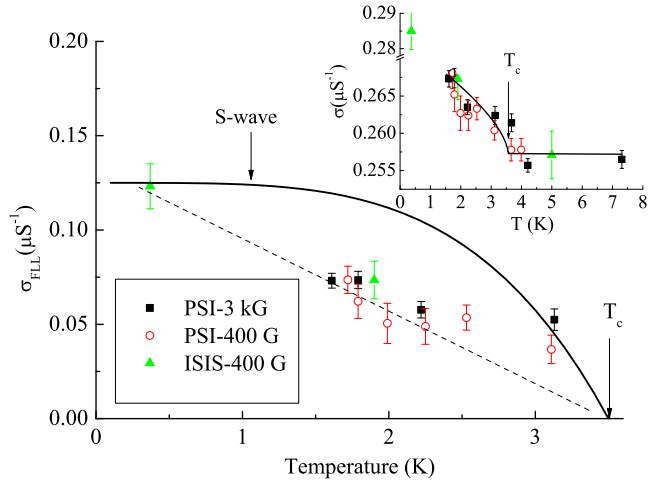}}

\caption{The part of the muon relaxation associated with the flux lattice
formation $\sigma_{FLL}$ as function of temperature. The solid line is the
temperature dependence predicted by the two-fluid model and an S-wave gap. The
dashed line is a guide to the eye. Inset: The total muon relaxation rate, as
taken in ISIS and PSI in a field of 400G and 3kG.}%
\label{Sigma}%

\end{figure}

However, for analysis purposes, in order not to degrade the data by the RRF
transformation, we fitted the 32 raw histograms separately. The \ same holds
for the GPS spectrometer which contains only $3$ counters in TF mode. We did
not group the counter histograms nor bin them in the fits. The fit function is
Gaussian since in a powder samples it describes the data sufficiently well.
Also this Gaussian is not sensitive to core radius and to the symmetry of the
flux lattice. To account for muon that missed the sample we used two Gaussian
relaxation functions one with very slow relaxation representing muons hitting
the Ti cell. The over all function is given by:
$A(t)=A_{0}\exp\left(  -\frac{(\sigma t)^{2}}{2}\right)  \cos(\gamma
Bt)+A_{cell}\exp\left(  -\frac{(\sigma_{cell}t)^{2}}{2}\right)  \cos(\gamma
B_{cell}t)$.

where $A_{0}$ and $A_{cell}$ are the initial asymmetries, $\sigma$ and
$\sigma_{cell}$ are the relaxation rates, and $B$ and $B_{cell}$ are the
averaged fields in the sample and cell respectively. The results of the fit
for the PSI data indicates that around $7\%$ of the muons missed the sample.
In ISIS the background signal is negligible due to the larger sample used. In
the inset of Fig.~\ref{Sigma} we show $\sigma$ vs. temperature for the data
taken in PSI in $400~$G and in $3~$kG, and in ISIS at $400$~G.%

As can be seen the change in relaxation in passing through $T_{c}$ is quite
small; between 4~K and 2~K it is only about $5\%$ of the normal state
relaxation $\sigma_{n}$. As mentioned before, $\sigma_{n}$ stems from nuclear
moments and seems to be field independent. Below $T_{c}$ the relaxation is
from a combination of nuclear moments and the flux lattice formed in the
sample. When the origin of the relaxation is a convolution of two
distributions, it results in a multiplication of two relaxation functions in
the time domain. Since both the nuclear moments and the flux lattice in a
powder sample generate Gaussian field distributions we can obtain the FLL part
by: $\sigma_{FLL}=\sqrt{\sigma^{2}-\sigma_{n}^{2}}$. In Fig.~\ref{Sigma} we
show $\sigma_{FLL}$ as function of temperature below $T_{c}$. This figure
includes all the data from PSI and ISIS, and the point at $T=0.37K$.
Unfortunately, due to experimental problems we do not have at present data
between $0.37K$ and $1.6K$. As one can see there is no field dependence, as
expected for a compound with $H_{c2}\sim61$~T \cite{Sakurai}. The penetration
depth $\lambda$ at base temperature is calculated from Eq.~\ref{deltaB} and
$\sigma_{FLL}=\gamma_{\mu}\sqrt{\left\langle \Delta B^{2}\right\rangle }$
where $\gamma_{\mu}=85.16MHz/kG$ is the gyromagnetic constant of the muon.
This calculation gives $\lambda=9100(500)\mathring{A}$ at $T=0.37$~K, which is
very large, and on the order of what is considered as the limit of TF-$\mu$SR.
Indeed, the error bars on $\sigma_{FLL}(T)$ are quite big and there is scatter
in the data. Nevertheless it is clear from Fig.~\ref{Sigma} that the
temperature dependence of $\sigma_{FLL}$ is inconsistent with the
phenomenological \textquotedblleft two-fluid\textquotedblright\ model
prediction: $\sigma(T)\propto1-(T/T_{c})^{4}$ \cite{Tinkham}. The fact that
$\sigma_{FLL}$ does not saturate even at low temperatures indicates that there
are nodes in the gap. Higemoto \textit{et. al.} \cite{musr_naco} reached a
similar conclusion based on muon Knight shift results which indicated a
non-complete gap.%

\begin{figure}
[ptb]
\centerline{\epsfxsize=9.5cm \epsfbox{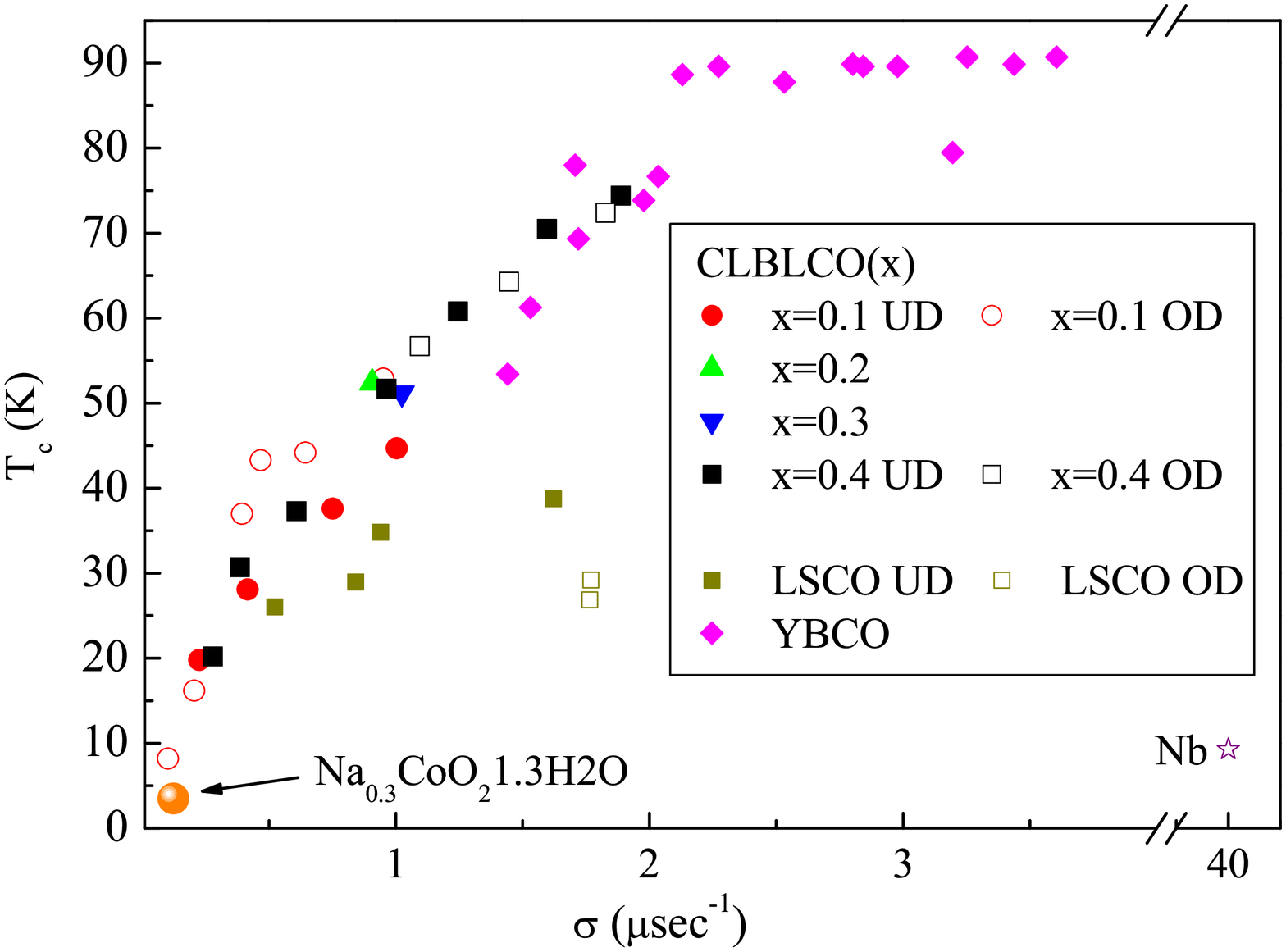}}

\caption{The Uemura plot showing $T_{c}$ vs. the muon relaxation rate $\sigma$
at the lowest temperature for LSCO, YBCO, and CLBLCO(x) cuprates, and for Nb.
The results from $Na_{x}CoO_{2}\cdot yH_{2}O$ fall on the line defined by the
cuuprates.}%
\label{Uemura}%

\end{figure}

Next, we would like to compare the superfluid density of the Na$_{0.3}%
$CoO$_{2}\cdot1.3$H$_{2}$O system with that of other unconventional SC and see
if it agrees with the Uemura relation. In Fig.~\ref{Uemura} we depict the
original Uemura line using data only from the high temperature superconductors
YBa$_{2}$Cu$_{3}$O$_{y}$ (YBCO) \cite{UemuraPRL91}, La$_{1-x}$Sr$_{x}$%
CuO$_{4}$ (LSCO) \cite{UemuraPRL91}, and (Ca$_{x}$La$_{1-x}$)(Ba$_{1.75-x}%
$La$_{0.25+x}$)Cu$_{3}$O$_{y}$ [CLBLCO(x)] \cite{KerenSSC03}. Underdoped and
overdoped samples are presented with solid and open symbols. The $\sigma
_{FLL}=0.125(10)$ at $T=0.37$~K and $T_{c}=3.5$~K fall exactly on this line.
For comparison we added the data for Nb, which is a BCS type II SC with
$T_{c}=9.26K$.

Finally we discuss the gender of Na$_{0.3}$CoO$_{2}\cdot1.3$H$_{2}$O. Assuming
$0.3$ free electrons per Co \cite{Cava_nature}, we get for our system a free
electron density of about $3.8\times10^{21}/cm^{3}$, which is comparable with
the value for optimally doped YBCO. The free electron density of $Nb$, for
example, is $5.56\times10^{22}/cm^{3}$, about an order of magnitude higher
than in Na$_{0.3}$CoO$_{2}\cdot1.3$H$_{2}$O. Using the London equation
\begin{equation}
1/\lambda^{2}=4\pi n_{s}e^{2}/m^{\ast}c^{2}%
\end{equation}
where $n_{s}$ is the superconducting carriers density and $m^{\ast}$ is the
effective mass of this carriers we can calculate the superfluid density. The
separation of $n_{s}/m^{\ast}$ is impossible using $\mu$SR alone and it is
very hard in general. But we can get useful insight from the comparison with
Nb. Assuming that roughly all the normal state carriers contribute to the
superconductivity, so that at $T\rightarrow0$ the superfluid density equals
the free electron density, we can extract the effective mass of the carrier
from $\lambda$. In the case of Nb we get an effective mass $m^{\ast}\sim
3m_{e}$, on the other hand for Na$_{0.3}$CoO$_{2}\cdot1.3$H$_{2}$O we get
$m^{\ast}\sim75m_{e}$. Despite this crude estimation, the mass is huge and
comparable to the effective masses of the heavy fermion superconductors. It is
much larger than the mass that the same calculation will yield for YBCO
($m^{\ast}\sim2m_{e}$), for example. In fact, thermopower \cite{Fisher} and
specific-heat measurements \cite{Cp,Cp2} point to the narrow band character of
the Na$_{x}$CoO$_{2}\cdot y$H$_{2}$O system which results in an enhancement of
the electron mass.%

In summary, we performed TF-$\mu$SR experiments on a sample of Na$_{0.3}%
$CoO$_{2}\cdot1.3$H$_{2}$O. The temperature dependence of the penetration
depth $\lambda$ indicates that superconductivity in this system is
unconventional and that the order parameter has nodes. The value of the
relaxation rate at low temperature agrees with the well known prediction of
the Uemura line. Comparing the normal state carrier density with the
super-fluid density reveals an unusually heavy superconductive carrier.

We would like to thank ISIS and PSI facilities for their kind hospitality. We
are grateful to A. Auerback for very helpful discussions. This work was funded
by the Israeli Science Foundation.

\end{document}